\documentclass[twocolumn,superscriptaddress,showpacs]{revtex4}
\usepackage{graphicx}

\begin{document}
\title{Diffractionless propagation through Kapitza stratified media}

\author{Carlo Rizza}
\affiliation{Dipartimento di Scienza e Alta Tecnologia, Universit\`a dell'Insubria, Via Valleggio 11, 22100 Como, Italy} \affiliation{Consiglio Nazionale
delle Ricerche, CNR-SPIN, 67100 Coppito L'Aquila, Italy}

\author{Alessandro Ciattoni}
\affiliation{Consiglio Nazionale delle Ricerche, CNR-SPIN, 67100 Coppito L'Aquila, Italy}

\begin{abstract}
We show that in the presence of a rapidly modulated dielectric permittivity with a large modulation depth (Kapitza medium) a novel and robust regime of
diffractionless electromagnetic propagation occurs. This happens when the mean value to depth ratio of the dielectric profile is comparable to the small
ratio between the modulation period and the wavelength. We show that the standard effective medium theory is inadequate to describe the proposed regime
and that, its occurrence is not substantially hampered by medium losses. We check the feasibility of the proposed regime by means of a large modulation
depth metal-dielectric layered medium whose subwavelength imaging properties are analytically investigated.
\end{abstract}

\maketitle

Subwavelength imaging and diffractionless propagation are major photonics subjects that have attracted, in the past decade, a renewed research interest
triggered by the advent of metamaterials \cite{Vesela,Pendr1}. In fact, negative index materials have been shown to support negative refraction of
propagating waves and, at the same time, magnification of the evanescent waves, thus providing an overall superlensing effect. Remarkably, without
resorting to exotic and hardly available magnetic properties, it has been realized that a metal slab can provide subwavelength imaging as well, the effect
being supported by plasmon excitation at its surfaces \cite{FangFa,Taubne}. A refinement of the metal slab superlens has been proposed in the form of
metal-dielectric layered composite obtained by alternating two materials of equal thickness and permittivities with real parts of opposite signs
\cite{Pendr2,Ramakr}. Basically, the chosen dielectric-metal symmetrical composition provides, within the standard effective medium approach, transverse
and longitudinal permittivities which are zero and infinite, respectively, thus effectively enabling the structure to support the electrostatic situation
where each wave is replicated both in amplitude and phase. However, due to the high impedance mismatch between such structure and vacuum, a different
layered metal-dielectric composite has been proposed which can be tailored to operate in the canalization regime where the effective transverse
permittivity is equal to $1$ while the longitudinal permittivity is very large \cite{Belov1,LiLiL1,LiLiL2,LiLiL3}. Such dielectric anisotropic response
provides the effective medium with a very flat equal frequency contour, this allowing the slab to transport nearly all the spatial harmonics of the
impinging image and, at the same time, to fulfill the Fabry-Perot resonance condition for enhancing transmission. Another subdiffraction imaging effect
exploits a different kind of anisotropic metamaterials whose ordinary and extraordinary permittivities have different signs (indefinite or hyperbolic
media) \cite{JacobJ,LiuLiu,LeeLee,CasseC}. The related hyperlensing effect is based on hyperbolicity of the extraordinary waves equifrequency contour
whose positive curvature entails negative refraction and whose unboundedness intrinsically allows propagation of the subwavelength field features.
Recently, it has been proposed that a plasma in the presence of a strong magnetic field is able to provide subdiffraction imaging for the electromagnetic
frequency well below the plasma frequency \cite{ZhangZ}, and also such configuration exploits the indefinite (anisotropic) response of the magnetized
plasma whose hyperbolic equifrequency contour is very flat at sufficiently low frequencies.

In this Letter we show that a novel regime of diffractionless propagation occurs for Transverse Magnetic (TM) waves propagating, in the long wavelength
regime, through rapidly modulated stratified media with a large dielectric modulation depth (Kapitza media). The considered electromagnetic situation is
conceptually equivalent to that of the mechanical inverted pendulum whose pivot point is subjected to high-frequency vertical oscillations (Kapitza
pendulum) \cite{Kapitz,Landau}, since these oscillations produce a rapidly varying contribution to the lagragian function with a large modulation depth.
The concept has been successfully extended and implemented in the general context of quantum \cite{Rahavv} and nonlinear physics \cite{Chaoss}, such as
for kink solitons in perturbed sine-Gordon models, with applications for long Josephson junctions \cite{Kivsha}, guiding-center solitons in perturbed
Nonlinear Schroedinger Equations, with applications in optical pulse propagation in fiber \cite{Guidi1,Guidi2}, in quadratic nonlinearities with
applications for spatial quadratic solitons \cite{Guidi3} and in "nonlinearity management" in both Bose-Einstein Condensates dynamics and nonlinear optics
\cite{Manag1,Manag2,Manag3,Manag4}. In all these cases, even though the system is not able to follow the rapid external oscillations, yet these are able
to affect the average system dynamics by means of additional slowly-varying potential contributions (e.g., the rapid external oscillations yield the
stabilization of the inverted state of the mechanical pendulum). In the electromagnetic situation we consider in this Letter, the rapid and deep
dielectric spatial modulation entails the diffractionless propagation of the averaged electromagnetic field.

The electromagnetic field amplitudes ${\bf E}= E_x(x,z) \hat{\bf e}_x + E_z(x,z) \hat{\bf e}_z$, ${\bf H}= H_y(x,z) \hat{\bf e}_y$ associated with
monochromatic TM waves satisfy Maxwell equations
\begin{eqnarray} \label{Maxwell}
-\partial_x E_z + \partial_z E_x &=& i\omega \mu_0 H_y, \nonumber \\
\partial_z H_y &=&  i\omega \epsilon_0 \epsilon_x E_x,  \nonumber \\
\partial_x H_y &=& -i\omega \epsilon_0 \epsilon_z E_z,
\end{eqnarray}
where $\epsilon = {\rm diag}[\epsilon_x,\epsilon_y,\epsilon_z]$ is the medium relative dielectric tensor and the time dependence $e^{-i\omega t}$ has been
assumed. Here we consider a specific medium periodically modulated along the $z$-axis whose dielectric permittivity $\epsilon_x = \epsilon_z = \epsilon$
admits the Fourier series expansion
\begin{equation} \label{ep}
\epsilon =\epsilon_m + \sum_{n \neq 0} \left( a_n + \frac{b_n}{\eta} \right) e^{in\frac{K}{\eta} z}
\end{equation}
where $\epsilon_m$ and $(a_n + b_n/\eta)$ are the Fourier coefficients whereas $2 \pi \eta /K$ is the spatial period. Here $\eta$ is a dimensionless
parameter we have introduced to explore the asymptotic electromagnetic behavior pertaining the limit $\eta \rightarrow 0$ where both the grating amplitude
and its spatial frequency are very large (assuming that $K \approx k_0 = \omega/c$). Note that the coefficients $a_n$ account for a contribution to the
modulation which is not large, and it is essential for allowing the model to avoid unrealistic features as a permittivity imaginary part rapidly
oscillating between large negative and positive values. Since electromagnetic propagation is here characterized by two very different scales (i.e. the
radiation wavelength and the permittivity modulation period), it is natural to let each electromagnetic field component to separately depend on $z$ and
$Z=z/\eta$ (multiscale technique where $Z$ is the fast coordinate \cite{Sander}) and to represent it as a Taylor expansion up to first order in $\eta$,
i.e.
\begin{eqnarray}
A(x,z,Z) = \left[\overline{A}^{(0)} (x,z) + \widetilde{A}^{(0)} (x,z,Z) \right]+ \nonumber \\
           \eta \left[ \overline{A}^{(1)} (x,z) + \widetilde{A}^{(1)} (x,z,Z) \right]
\end{eqnarray}
where $A=E_x,E_z$ or $H_y$, the superscripts $(0)$ and $(1)$ label the order of each term whereas the overline and the tilde label the averaged and
rapidly varying contributions to each order, respectively. Substituting the field components of this form into Eqs.(\ref{Maxwell}), each equation yields a
power series in $\eta$ whose various orders are evidently the superposition of a slowly varying (independent on $Z$) and a fast (dependent on $Z$)
contribution all of which can independently be balanced. From the lowest order (which is here $\eta^{-1}$) we obtain $\widetilde{E}^{(0)}_x = 0$,
$\overline{E}^{(0)}_z = 0$, $\widetilde{E}^{(0)}_z = 0$ and $\widetilde{H}^{(0)}_y =(\omega \epsilon_0 /K) \sum_{n \neq 0} (b^{(n)}/n ) e^{inKZ}
\widetilde{E}^{(0)}_x$ or, in other words, the dominant contribution to $x$-component of electric field is slowly varying, whereas the $z$-component of
the electric field has no zeroth order. The order $\eta^0$ of the first of Eqs.(\ref{Maxwell}) yields
\begin{equation} \label{Max1}
\partial_z \overline{E}^{(0)}_x = i \omega \mu_0 \overline{H}^{(0)}_y
\end{equation}
and $\partial_Z \widetilde{E}^{(1)}_x = i\omega \mu_0 \widetilde{H}^{(0)}_y$, which in turn, with the help of the obtained expression for
$\widetilde{H}^{(0)}_y$, yields $\widetilde{E}^{(1)}_x = (k_0^2/K^2) \sum_{n \neq 0} (b^{(n)}/n^2 ) e^{inKZ} \widetilde{E}^{(0)}_x$. The slowly varying
part of the order $\eta^0$ of the second of Eqs.(\ref{Maxwell}), together with the obtained expression for $\widetilde{E}^{(1)}_x$ yield
\begin{equation} \label{Max2}
\partial_z \overline{H}^{(0)}_y = -i \omega \epsilon_0 \epsilon_{eff} \overline{E}^{(0)}_x.
\end{equation}
where
\begin{equation} \label{epeff}
\epsilon_{eff} = \epsilon_m + \frac{k_0^2}{K^2} \sum_{n \neq 0} \frac{ b^{(-n)} b^{(n)} }{n^2}.
\end{equation}
To summarize we have obtained that the leading term of the TM electromagnetic field is {\it slowly varying} and rigorously Transverse Electro-Magnetic
(TEM), i.e. ${\bf E} \simeq \overline{E}^{(0)}_x(x,z) \hat{\bf e}_x$, ${\bf H} \simeq \overline{H}^{(0)}_y (x,z) \hat{\bf e}_y$, and, in addition, its
components satisfy Eqs.(\ref{Max1}) and (\ref{Max2}) so that the field {\it rigorously} undergoes diffractionless propagation and experiences the uniform
effective permittivity $\epsilon_{eff}$.

As for any Effective Medium Theory (EMT), the field is here slowly varying since it cannot follow the rapid medium dielectric modulation. However, the
fact that the longitudinal component vanishes is a very unique and important feature of the proposed regime which has a simple physical explanation
arising from the combination of the first Maxwell Equation and the large depth of the dielectric modulation. In fact, a very deep and rapidly modulated
dielectric response would produce volume polarization charges too strong to be screened by the electric field, an incompatibility avoided by the condition
$E_z \simeq 0$. For example, assuming $\epsilon = \epsilon_m + (b/\eta) \cos (Kz/\eta)$, the equation $\nabla \cdot {\bf D} =0$ yields $\epsilon
(\partial_x E_x + \partial_z E_z) - (K/\eta^2) \sin (Kz/\eta) E_z = 0$ from which it is evident that, since the first term self-consistently scales as
$1/\eta$, $E_z$ has to be proportional to $\eta$ for mitigating the divergence of the factor $1/\eta^2$ in the second term. The diffractionless
propagation of the leading contribution to the electromagnetic field in the considered regime is a consequence of its TEM structure since, as it is well
known, TM waves with vanishing longitudinal component do not undergo diffraction (as it is evident from Eqs.(\ref{Maxwell})). It should be stressed that
the super-resolution mechanisms reported in literature are based on the requirement that $\epsilon_z \rightarrow \infty$ (since, when combined with the
third of Eqs.(\ref{Maxwell}), this condition yields $E_z \rightarrow 0$ and hence diffractionless propagation) a condition that has to be fulfilled by
exploiting the standard EMT and which is essentially hampered by medium absorption \cite{LiLiL1}. On the other hand, in the regime we are investigating
here, the longitudinal effective permittivity $\epsilon_z$ is not even defined and it is evident from the above presented multiscale analysis that the
longitudinal field component $E_z$ vanishes regardless the medium losses (since we have not assumed the permittivity to be real). We conclude that the
occurrence of diffractionless propagation within Kapitza stratified media is very robust and substantially not hampered by medium losses. Note that, in
analogy to the potential governing the averaged dynamics of the Kapitza pendulum, the effective permittivity of Eq.(\ref{epeff}) is the sum of the average
permittivity plus a contribution arising from the rapidly varying part of the dielectric modulation. Such an effective medium response, together with the
TEM field structure, clarifies that the Effective Medium Theory we are considering here (Kapitza EMT) is fundamentally different from the Standard
Effective Medium theory (Standard EMT) for layered media whose main result is that the effective principal permittivities are $\epsilon_x = \langle
\epsilon \rangle = \epsilon_m$ and $\epsilon_z = \langle \epsilon^{-1} \rangle ^{-1}$. The difference between the two regimes is also assured by the fact
that the standard effective medium results can be simply derived by using the above discussed multiscale technique in the presence of the dielectric
profile of Eq.(\ref{ep}) with $b_n =0$, i.e. without the large modulation depth contribution to the permittivity.

\begin{figure}
\includegraphics[width=0.45\textwidth]{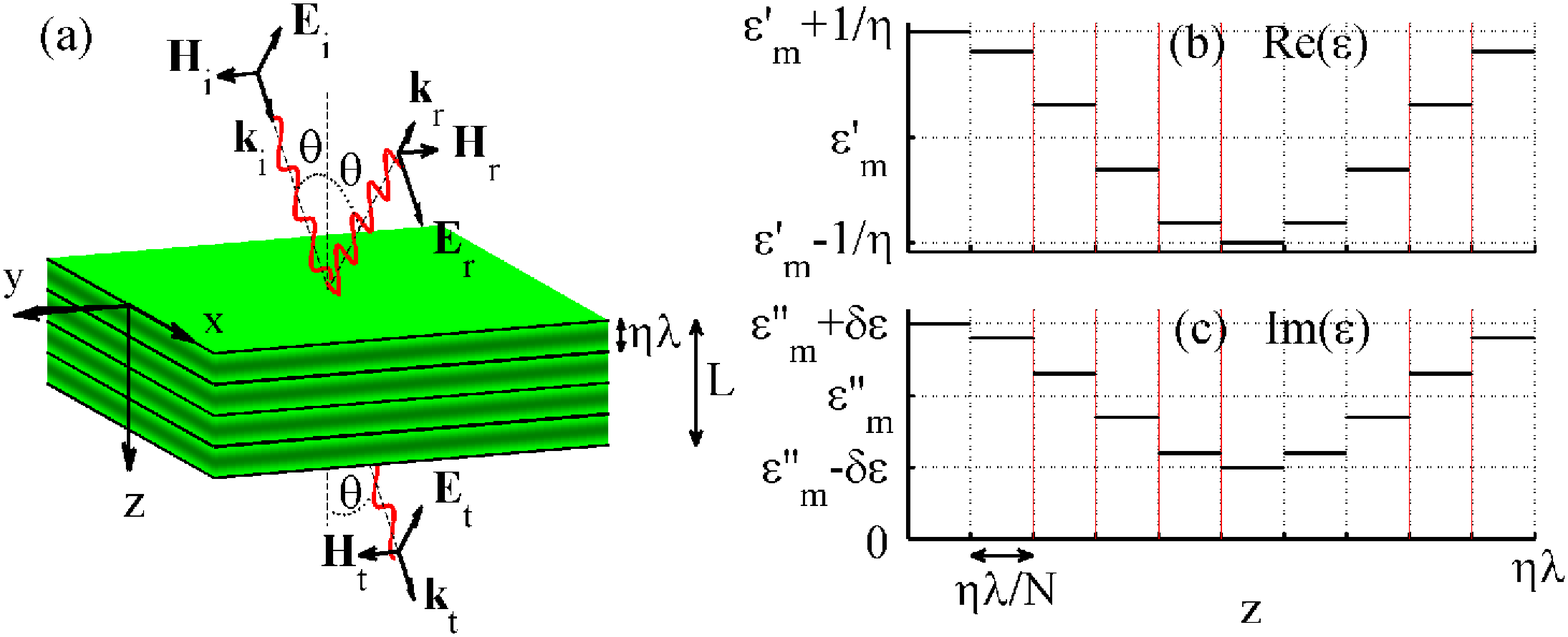}
\includegraphics[width=0.45\textwidth]{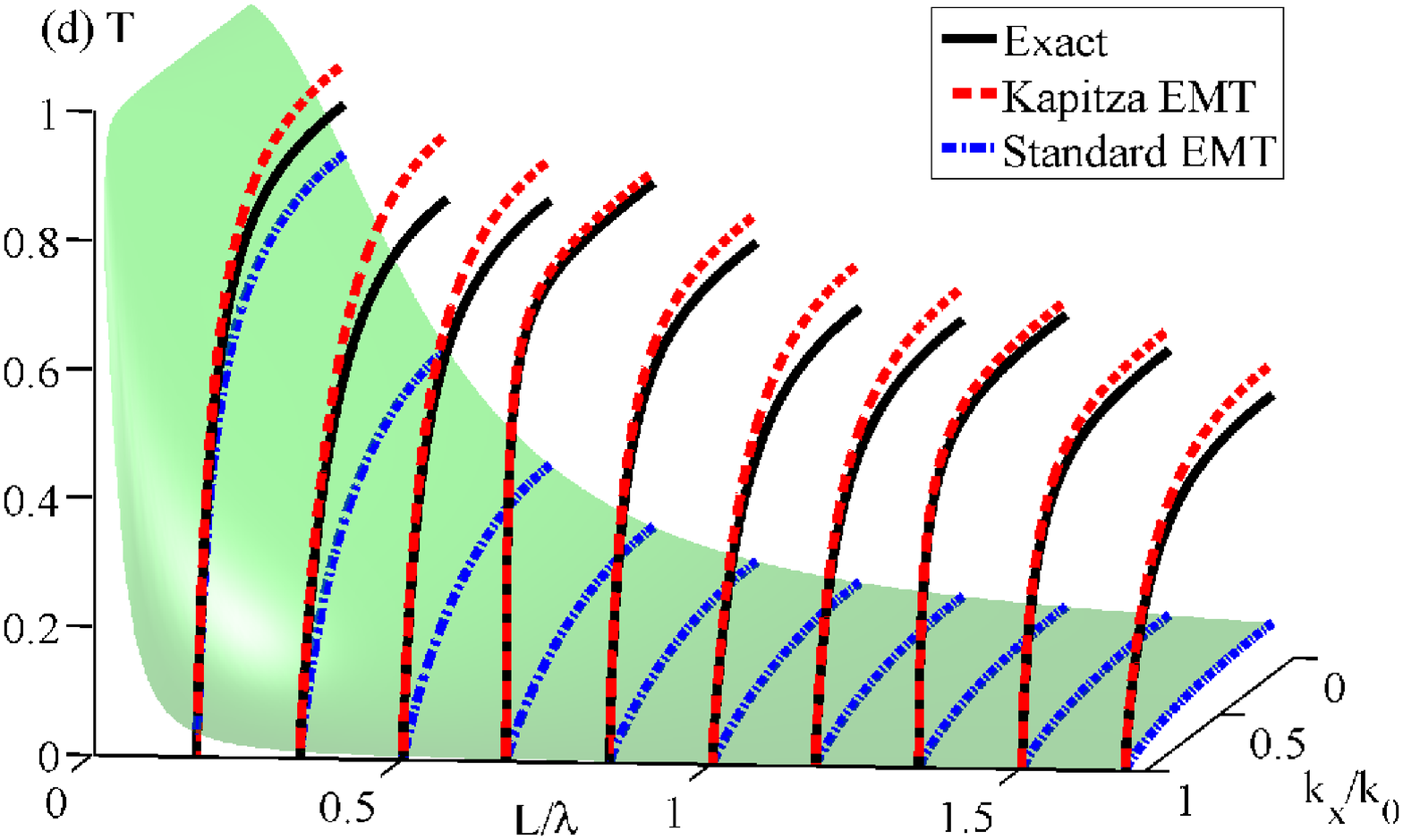}
\caption{(Color online) (a) Layered slab and TM waves scattering geometry. (b) and (c) real and imaginary parts of the dielectric permittivity profile
within the unit cell. (d) Comparison between the exact slab transmissivity (solid lines) and thoose predicted by the Kapitza (dashed lines) and the
standard EMTs (dash dotted lines) for a the unit cell dielectric distribution of panel (b) and (c) with $\epsilon_m = 0.05 +0.05 i$, $\delta \epsilon =
0.025$, $N=10$ and $\eta = 1/ 60$, and wavelength $\lambda = 100 \: \: \mu m$. The semitransparent surface interpolating the standard EMT transmissivity
profiles is plotted for clarity purposes.}
\end{figure}

In order to check the predictions of the proposed Kapitza EMT, we have considered reflection and transmission of TM plane waves from a slab filled by a
Kapitza stratified medium, as sketched in Fig.1(a). The dielectric modulation is along the $z$-axis with period $\eta \lambda$ where $\lambda = 2\pi /
k_0$ is the vacuum radiation wavelength and $\eta$ is the above introduced small parameter, whereas the slab thickness $L$ is a multiple of the period
$\eta \lambda$. The unit cell comprises $N$ homogeneous layers of thicknesses $\eta \lambda / N$ and the dielectric permittivity of the $j-th$ layer
($j=1,...,N$) is
\begin{equation} \label{epL}
\epsilon_j = \epsilon_m + \left( \frac{1}{\eta} + i \delta \epsilon \right) \cos \left[ \frac{2\pi}{N} (j-1) \right]
\end{equation}
where $\epsilon_m$ is the mean value of the dielectric permittivity and $\delta \epsilon$ is responsible for the (not large) modulation of medium
absorption. The corresponding profile of the slab dielectric permittivity is evidently of the kind of Eq.(\ref{ep}) (with $K=k_0$) and its real and
imaginary parts are sketched, within a unit cell, in Figs.1(b) and 1(c), respectively (where $\epsilon'_m={\rm Re}(\epsilon_m)$ and $\epsilon''_m={\rm
Im}(\epsilon_m)$). The check of the Kapitsa EMT has been performed by choosing $\lambda = 100 \: \mu m$, $\epsilon_m = 0.05 +0.05 i$, $\delta \epsilon =
0.025$, $N=10$ and $\eta = 1/ 60$, which is a realistic situation since $-60 < {\rm Re} (\epsilon) < 60$ and $0.025 < {\rm Im} (\epsilon) < 0.075$ (thus
avoiding the use of active media) and since the layers' thickness $\eta \lambda / N \simeq 166 \: \: nm$ is fully feasible. The scattering process of TM
waves by the considered layered medium admits full analytical description (by means of the transfer-matrix method) and in Fig.1(d) we have plotted the
exact profiles (solid line) of the transmissivity defined as $T = |{\bf E}_t|^2 / |{\bf E}_i|^2$ (see Fig.1(a) for the definition of the field amplitudes)
as function of the transverse wave vector $k_x = k_0 \sin \theta$ for various slab thicknesses $L$. On the other hand, Eqs.(\ref{Max1}) and (\ref{Max2})
are easily solved to yield the transmissivity
\begin{equation} \label{TK}
T  =  \left| \cos(k_z L) - i F \sin (k_z L) \right|^{-2}
\end{equation}
where $k_z = k_0 \sqrt{ \varepsilon_{eff}}$ and $F=\left[\sqrt{ \varepsilon_{eff}} \cos \theta + 1/ \left(\sqrt{ \varepsilon_{eff}} \cos \theta
\right)\right]/2$. After evaluating the Fourier coefficients of the considered dielectric profile, Eq.(\ref{epeff}) yields $\epsilon_{eff} = 0.5339 +
0.05i$ and in Fig.1(d) we have reported various profiles (dashed lines) of the resulting transmissivity of Eq.(\ref{TK}) pertaining to the Kapitza EMT. In
addition, the standard EMT would describe the slab as an anisotropic medium with dielectric permittivities $\epsilon_x = \langle \epsilon \rangle =
0.05+0.05i$ and $\epsilon_z = \langle \epsilon^{-1} \rangle ^{-1} = (-7.205 + 7.194i) \cdot 10^3$ and, in Fig.1(d), we have plotted various profiles
(dashed dot lines) of the the corresponding transmissivity which coincides with Eq.(\ref{TK}) with $k_z = k_0 \sqrt{ \epsilon_x \left(1 -
\frac{\sin^2\theta}{\epsilon_z} \right)}$ and $F = \frac{1}{2} \left(\frac{k_0 \epsilon_x \cos \theta}{k_z} + \frac{k_z}{k_0 \epsilon_x \cos
\theta}\right)$. Note that the agreement between the exact predictions and those based on the Kapitza EMT is remarkable whereas the standard EMT
completely fails.

\begin{figure}
\includegraphics[width=0.45\textwidth]{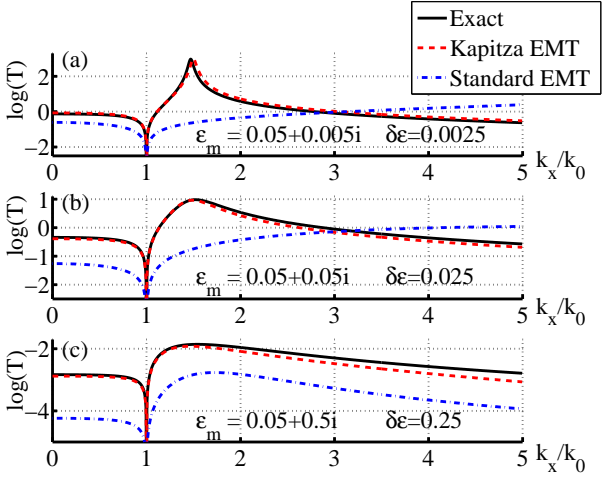}
\caption{(Color online) Logarithmic plot of the quantity $T = |{\bf E}_t|^2_{z=L} / |{\bf E}_i|^2_{z=0}$, as function of the transverse impinging momentum
$k_x$, generalizing the slab transmissivity of Fig.1(d) to encompass transport of evanescent waves for three slabs of thickness $L =1.66 \lambda$ and
$\eta = 1/60$, with different mean values and modulation depths of the permittivity imaginary part.}
\end{figure}

It is well known that medium absorption plays a very detrimental role in the observation of the canalization regime \cite{LiLiL1} as well as on all the
other mechanisms supporting diffractionless propagation. In order to investigate the effect of medium losses on the Kapitza regime we are here discussing,
we have extended the scattering experiment of Fig.1(a) to encompass even evanescent waves and we have evaluated, for a number of slabs characterized by
different absorption efficiencies, the quantity $T = |{\bf E}_t|^2_{z=L} / |{\bf E}_i|^2_{z=0}$ as a function of the transverse wave vector $k_x$ (which,
for $|k_x| < k_0$, coincides with the above discussed slab transmissivity and, for $|k_x| > k_0$, entails information on the slab efficiency to transport
evanescent waves). In Fig.2 we report the exact logarithmic plot of $T$ together with the corresponding profiles predicted by the Kapitza and the standard
EMTs for three different slabs of thickness $L=1.66 \lambda$ which are identical to those considered in Fig.1 but characterized by different values of
$\epsilon_m$ and $\delta \epsilon$: (a) $\epsilon_m = 0.05 + 0.005 i$, $\delta \epsilon = 0.0025$; (b) $\epsilon_m = 0.05 + 0.05 i$, $\delta \epsilon =
0.025$; (c) $\epsilon_m = 0.05 + 0.5 i$, $\delta \epsilon = 0.025$. The corresponding dielectric permittivities predicted by the Kapitza and the standard
EMTs are: (a) $\epsilon_{eff} = 0.5339+0.0050i$, $\epsilon_x=0.05+0.005i$, $\epsilon_z = (-14.256+1.424i) \cdot 10^3$; (b) $\epsilon_{eff} = 0.5339 +
0.05i$, $\epsilon_x = 0.05+0.05i$, $\epsilon_z = (-7.205 + 7.194i) \cdot 10^3$; (c) $\epsilon_{eff} = 0.5339+0.05i$, $\epsilon_x=0.05+0.5i$, $\epsilon_z =
(-0.154+1.425i) \cdot 10^3$. From Fig.2 we note that in all the three considered situations the Kapitza EMT predictions largely agree with the exact
phenomenology whereas the standard EMT is inadequate and, most importantly, that this happens almost regardless the medium absorption.Considering that in
panel (c) of Fig.2 the mean value and the modulation depth of the permittivity imaginary part are one hundred times greater than those of panel (a), we
restate that the Kapitza regime is very robust against medium losses, a very distinct and important feature which does not characterize the other
diffractionless propagation regime presented in literature.

\begin{figure}
\includegraphics[width=0.45\textwidth]{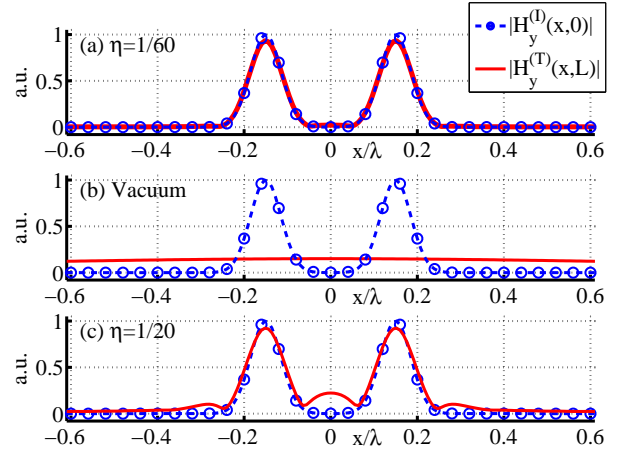}
\caption{(Color online) (a) Imaging of a subwavelength two-peaked shape at $\lambda = 100 \: \: \mu m$ by a Kapitza slab (as reported in Fig.1(a) with
unit cell dielectric profile of Figs. 1(b) and 1(c)) characterized by $\epsilon_m = 0.05 + 0.005 i$, $\delta \epsilon = 0.0025$, $N=10$, $\eta = 1/60$ and
$L=135.66 \: \: \mu m$. (b) Diffraction of the subwavelength two-peaked shape after $L=135.66 \: \: \mu m$ vacuum propagation. (c) Same as in panel (a)
but with $\eta = 1/20$.}
\end{figure}

From Fig.2, it should also be stressed that the agreement between the Kapitza EMT and the exact phenomenology is, in the considered situation, very good
and robust even for evanescent waves ($|k_x|> k_0$), an essential requirement for achieving super-resolution. Even though we have mainly focused on the
novel diffractionless propagation regime, it is worth checking the subwavelength imaging features of a Kapitza layered slab. Specifically we have
considered a slab whose unit cell has the dielectric distribution of Eq.(\ref{epL}) with $\epsilon_m = 0.05 + 0.005 i$, $\delta \epsilon = 0.0025$, $N=10$
and $\eta = 1/60$ (thus coinciding with the case of Fig.2(a)). The radiation wavelength is $\lambda=100 \: \: \mu m$ and the unit cell thickness is
$\Lambda = \eta \lambda = 1.67 \: \: \mu m$. We have considered a slab with $M=82$ unit cells since its resulting width $L= M\Lambda = 135.66 \: \: \mu m$
is very close to the width $L_F = \lambda / \sqrt{{\rm Re}(\epsilon_{eff})} = 136.85 \: \: \mu m$ for which the Kapitza homogeneous slab governed by
Eqs.(\ref{Max1}) and (\ref{Max2}) with $\epsilon_{eff}=0.5339+0.005i$ (from Eq.(\ref{epeff})) has its second Fabry-Perot resonance. Using the
transfer-matrix method, we have evaluated the slab point-spread-function $PSF(x)$ according to which the transmitted magnetic field profile
$H_y^{(T)}(x,L)$ can be expressed as the convolution of $PSF(x)$ with the {\it incident} magnetic field $H_y^{(I)}(x,0)$, or $H_y^{(T)}(x,L) =
PSF(x)*H_y^{(I)}(x,0)$ \cite{Kotyns}. In Fig.3(a) we plot the comparison between the slab output and input images for an impinging magnetic field whose
profile consists of two Gaussians of equal width $\sigma = \lambda/20$ whose separation distance is $4 \sigma$, i.e. $H_y^{(I)}(x,0) = H_0 \left[
e^{-(x-2\sigma)^2/\sigma^2} + e^{-(x+2\sigma)^2/\sigma^2}\right]$. Apart from a slight decrease of the peaks height due to medium absorption, the
outcoming image is clearly a very accurate copy of the incoming one. To appreciate such imaging result, we have reported in Fig.3(b) the effect of vacuum
propagation on the above incoming field, and the complete deterioration of the considered subwavelength image is evident. For completeness, in Fig.3(c) we
have plotted the analogous fields comparison for a slab identical to that of panel (c) but with $\eta = 1/20$ whose subwavelength imaging efficiency is
evident even in the present of a slight image deterioration resulting from the small sidelobes.

In conclusion, we have proposed and discussed a novel diffractionless propagation regime which is not fundamentally affected by medium losses and it is
very robust against the tailoring of the medium dielectric properties. Diffraction suppression is here achieved through propagation in a medium whose
dielectric permittivity is rapidly modulated with a large modulation depth so that the physical underpinnings of the proposed setup are in many respects
an elaboration of those that in mechanics are associated with the stabilization of the inverted pendulum (the Kapitza pendulum from which the name Kapitza
EMT for our approach). In addition to the broad interest of our results (obtained from the physical ubiquitous situation where a rapidly varying and
strong external stimulus is able to affect the slow system dynamics) we stress that our method is amenable to many relevant elaborations, as for example,
the extension to the case where also the medium magnetic permeability is rapidly varying with a large modulation dept. We suggest that such a situation,
surpassing the here considered TM electromagnetic case, could in principle be able to support diffractionless propagation of the full electromagnetic
field.

This research has been funded by the Italian Ministry of Research (MIUR) through the "Futuro in Ricerca" FIRB-grant PHOCOS - RBFR08E7VA.


\end{document}